\documentclass[prd,twocolumn,showpacs,amsmath,amssymb,floatfix,nofootinbib]{revtex4}
\usepackage{graphicx,dcolumn,booktabs}
\usepackage{amsmath}
\usepackage{amsfonts}
\usepackage{amssymb}
\usepackage{times}
\usepackage{amsmath, amssymb}

\begin{document}
\title{Phenomenological study on the significance of the
 scalar potential and Lamb shift}

\author{ Xu-Hao~Yuan $^{1}$   \footnote{segoat@mail.nankai.edu.cn},
         Hong-Wei~Ke $^{2}$  \footnote{khw020056@hotmail.com}
         Xue-Qian~Li $^{1}$  \footnote{lixq@nankai.edu.cn} }

\affiliation{
  $^{1}$ School of Physics, Nankai University, Tianjin 300071,
  China\\
  $^{2}$ School of Science, Tianjin University, Tianjin 300072, China }

\begin{abstract}
\noindent We indicated in our previous work that for QED the
contributions of the  scalar potential which appears at the loop
level is much smaller than that of the vector potential and in fact
negligible. But the situation may be different for QCD, one reason
is that the loop effects are more significant because $\alpha_s$ is
much larger than $\alpha$, and secondly the non-perturbative QCD
effects may induce the scalar potential. In this work, we
phenomenologically study the contribution of the scalar potential to
the spectra of charmonia. Taking into account both vector and scalar
potentials, by fitting the well measured charmonia spectra, we
re-fix the relevant parameters and test them by calculating other
states of the charmonia family. We also consider the role of the
Lamb shift and present the numerical results with and without
involving the Lamb shift.
\end{abstract}

\maketitle

\section{Introduction}

By a symmetry consideration, Chen et al. suggested \cite{chen:2008}
that for the Coulomb interaction, to maintain the hidden symmetry
SO(4) in the Schr\"odingger equation, the scalar and vector
potentials must have the same weight in the Dirac equation. The
hidden symmetry is just the familiar Lenz symmetry which also exists
in the classical physics. However, if so, the orbit-spin coupling
would disappear. In fact the scalar and vector potentials make
opposite contributions to the orbit-spin coupling, thus if they have
the same weight, their contributions would exactly cancel each
other. It definitely contradicts to the data. Therefore, one
concludes that this symmetry does not exist in the relativistic
extension. Usually, one is tempted to think that the relativistic
Dirac equation should possess a higher symmetry than its
non-relativistic approximation, but this is not the case we are
confronting. A general theory which only considers the Lorentz
structure of the vertices\cite{Lucha:1991it}, there are five types
of coupling. But which one dominates should be selected by the
underlying physics. We turn to look at the deeper side, namely start
to investigate the problem in the quantum field theory.

The basic theory which induces the electric Coulomb potential is QED
whose coupling is vector-type $\bar\psi\gamma_\mu\psi A^{\mu}$, thus
at the tree-level, the induced potential is the vector one and the
other types should be induced at higher order, i.e. loop level. In
our earlier work \cite{Ke:2009mx}, we showed explicitly that the
scalar coupling $1\otimes 1$ which results in the scalar potential,
appears at the loop level and its contribution is suppressed by a
factor $\alpha/\pi$. For QED it is a small value and cannot make a
sizable contribution. Thus the apparent SO(4) symmetry at the
classical level is almost fully violated. However, the situation
would be different for the QCD case, because first $\alpha_s$ at the
charm-mass-scale is much larger than $\alpha$ and secondly the
non-perturbative QCD effects may also cause the scalar potential.

This case is noticed by Leviatan and some studies have been carried
out \cite{Leviatan:2003wy,Leviatan:2004mr,Leviatan:2009rg}. In this
work, we are not going to further discuss the origin of the scalar
potential or try to derive it from the quantum field theory, but
generally assuming its existence and by fitting the spectra of the
charmonia family, we obtain its fraction. Moreover, the QED theory
predicts the Lamb shift which is due to the vacuum effects. In QM,
it only shifts the S-wave spectra because in the non-relativistic
limit, it is proportional to $\delta({\bf r})$, but by the quantum
field theory, the other $l$-states are also affected. In other
words, by considering the Lamb shift, the positions of the spectra
would deviate from that obtained without the Lamb shift. In this
work, we include its contribution and re-fit the charmonia spectra
to obtain a new set of the model parameters. For a comparison, we
will present the numerical results with and without taking  the Lamb
shift into account.

Unlike the hydrogen-like atoms where the nucleus is very heavy and
approximated at rest, therefore only the motion of electron is
considered and the corresponding equation, either the relativistic
Dirac equation or non-relativistic Schr\"odinger equation, is a
one-body equation. However, for charmonia, the charm and anti-charm
quarks are of the same mass and the equation which properly
describes charmonia, should be a two-body equation.

For simplicity but without losing the significant characters, we do
not directly solve the two-body Dirac equation which is very
complicated. One can derive the effective potential between the two
constituents ($c$ and $\bar c$) in terms of the perturbative theory
where the effective Lorentz vertices are set according to the
general Lorentz structures \cite{Lucha:1991vn}. Because of the
limitation of the perturbative theory, we can only obtain the
Coulomb-type interaction and the corresponding spin-orbit,
spin-tensor and relativistic correction pieces. It is noted that the
fundamental QCD indeed provides only the vector potential at the
tree level, but as indicated above, the loop effect and even
non-perturbative effect may result in scalar potential. Thus we just
keep the potential forms and introduce two phenomenological
constants in front of the scalar and vector potentials and the
induced terms. For the confinement piece, we employ the linear
confinement i.e. the Cornell-type. In fact, the exact form of the
full potential including both scalar and vector pieces was given by
Lucha et al. \cite{Lucha:1991vn} and we just re-check their results
and then substitute the potential into our Schr\"odinger equation.

Now we can reduce the two-body Schr\"odinger equation into one
particle equation where the kinetic term is ${1\over 2\mu}{\bf p}^2$
where $\mu$ is the reduced mass and is $m_c/2$ in our case. Solving
the differential equation, we obtain the spectra. Since there exist
several phenomenological parameters which so far cannot be derived
from the underlying theory, we can  fix them by fitting a few well
measured charmonia states.

Moreover, as well known, the vacuum fluctuation induces the Lamb
shift. The basic Lagrangian of the Lamb shift has been derived by
some authors, and for interaction, we have $H_{int}=-L_{int}$
\cite{Hoang:2001rr,Titard:1993nn}. Thus we substitute the expression
into our data fitting process to re-derive the phenomenological
parameters. Indeed, the Lamb shift only occurs at the loop level,
but the Coulomb-type $-\alpha_s/r$ appears at the tree level of QCD.
It seems that they belong to different levels, but as we introduce
the phenomenological parameters which include the loop and
non-perturbative QCD effects, we cannot distinguish between the tree
level contribution and that of higher orders . However, for the Lamb
shift, we do not introduce a new phenomenological parameter, but use
the derived form \footnote{It is noted that for a formula which is
derived in the field theory, one can separate the contributions
corresponding to different orders as long as there is no
phenomenological parameters involved, and that is the case we deal
with the Lamb shift, please see the text for details}.

There are some subtleties in the calculations which we will address
in the text.

This paper is organized as follows. In Section \ref{sec-2} and
\ref{sec-3}, we introduce the generalized Breit-Fermi Hamiltonian
and the Sch$\ddot{\textrm{o}}$rdinger equation for the $c\bar{c}$
bound states: $J/\psi$, $\chi_{c0}(1\mathrm P)$, $\chi_{c1}(1\mathrm
P)$, $\eta_c(2\mathrm S)$ and $\psi(2\mathrm S)$. Then we
numerically solve the eigen-equations for these bound states and fix
the parameters. In Section \ref{sec-4}, the Lamb shift is concerned
and another set of the parameters is given to improve our
predictions. The last section is devoted to our conclusion and
discussion.

\section{The Generalized Breit-Fermi Hamiltonian and Sch\"ordinger
 equation}\label{sec-2}
For the $c\bar{c}$ meson, the generalized Breit-Fermi Hamiltonian
was given in Refs. \cite{Lucha:1991it,Ding:1991} as
\begin{subequations}\label{B-F Hamilton}
\begin{eqnarray}
 &&H=H_0+H_1+...,\\
 &&H_0={p^2\over m}+S(r)+V(r),\\
 &&H_1=H_{sd}+H_{si},\\
 &&H_{sd}=H_{ls}+H_{ss}+H_{t}\nonumber\\
 &&={1\over2m^2r}\left(3V'-S'\right)
 \vec{L}\cdot(\vec{S}_1+\vec{S}_2)+{2\over3m^2}\vec{S}_1\cdot\vec{S}_2\nabla^2V(r)\nonumber\\
 &&+{1\over12m^2}\left({1\over
 r}V'-V''\right)S_{12},
\end{eqnarray}
\begin{eqnarray}
 H_{si}&=&-{p^4\over4m^3}+{1\over4m^2}\left\{{2\over r}V'(r)\cdot \vec{L}^2
 +[p^2,V-r V']\right.\nonumber\\
 & &+\left.2(V-r V')p^2+{1\over2}\left[{8\over
 r}V'(r)+V''-r V'''\right]\right\}\nonumber\\
\end{eqnarray}
\end{subequations}
where, V and S stand for the vector and scalar potentials and
$H_{si}$ and $H_{sd}$ represent the spin-independent and
spin-dependent pieces respectively. For the linear confinement piece
we adopt the Cornell potential\cite{Eichten:cornell}. Thus the total
potential at lowest order reads
\begin{subequations}\label{v-s}
\begin{eqnarray}
 U(r)=V(r)+S(r)=-aC_F{\alpha_s\over r}+b\kappa^2r
\end{eqnarray}
\mbox{where},
\begin{eqnarray}
 \left\{
  \begin{array}{l}
   V(r)=-c~C_F\alpha_s/r+d\kappa^2r\\
   S(r)=-(a-c)C_F\alpha_s/r+(b-d)\kappa^2r
  \end{array}
 \right.
\end{eqnarray}
\end{subequations}
With the Hamiltonian (\ref{B-F Hamilton}) and the potential
(\ref{v-s}), one can solve the Sch\"ordinger equation
\begin{eqnarray}\label{sch.}
 (E-2m)\Psi(r)=H\Psi(r)=(H_0+H_1)\Psi(r).
\end{eqnarray}

If we define the radial wave function as $R(x)$ with the
dimensionless variable: $x=\kappa r$, then the radial equation is
written as\footnote{The standard form of the radial equation can be
easily found in \cite{Cai:2003}, and the method to make it
dimensionless is given in \cite{Silbar:2010}.}
\begin{subequations}\label{radl.eqn}
\begin{eqnarray}
 {d^2\over dx^2}u(x)=A(x)u(x)
\end{eqnarray}
where,
\begin{eqnarray}
 A(x)&=&-\tilde{m}\left(\tilde{E}-2\tilde{m}-\tilde{U}(x)-\tilde{H}_1\right)
  +{l(l+1)\over x^2}\nonumber\\
 & &-{1\over4}\left(\tilde{E}-2\tilde{m}-\tilde{U}(x)\right)^2
\end{eqnarray}
with
\begin{eqnarray}
 \left\{
  \begin{array}{l}
   \tilde{m}=m/\kappa,\quad \tilde{E}=E/\kappa,\\
   \tilde{H}_1=H_1/\kappa,~ \tilde{U}(x)=U(x)/\kappa.\\
  \end{array}
 \right.
\end{eqnarray}
The approximation
\begin{eqnarray}
 p^2\thickapprox m\left(E-2m-U(r)\right)
\end{eqnarray}
\end{subequations}
is used in (\ref{sch.}).

\section{The Energy Gap Function Of The $c\bar{c}$ charmonia and The numerical
Results}\label{sec-3}

The radial equation (\ref{radl.eqn}) can be solved in terms of the
method called `` the iterative numerical process `` which is
introduced in literatures, (for example, see
\cite{Silbar:2010,Cai:2003}). We have improved this method, and then
fix the parameters a, b, c, d by fitting the well measured spectra
of $c\bar{c}$ charmonia: $J/\psi$, $\chi_{c0}(1\mathrm P)$,
$\chi_{c1}(1\mathrm P)$, $\eta_c(2\mathrm S)$ and $\psi(2\mathrm
S)$. Instead of directly fitting the masses, we construct a series
of relations which should be fitted:

\begin{eqnarray}\label{e-g}
 \left\{
  \begin{array}{l}
   m\left[\psi(2\mathrm{S})\right]-m\left[\chi_{c1}(1\mathrm P)\right]
    =E\left[2^3\mathrm S_1\right]-E\left[1^3\mathrm P_1\right];\\
   m\left[\psi(2\mathrm S)\right]-m\left[J/\psi(1\mathrm S)\right]
    =E\left[2^3\mathrm S_1\right]-E\left[1^3\mathrm S_1\right];\\
   m\left[\psi(2\mathrm S)\right]-m\left[\eta(2\mathrm S)\right]
    =E\left[2^3\mathrm S_1\right]-E\left[2^1\mathrm S_0\right];\\
   m\left[\psi(2\mathrm S)\right]-m\left[\chi_{c0}(1\mathrm P)\right]
    =E\left[2^3\mathrm S_1\right]-E\left[1^3\mathrm P_0\right].
  \end{array}
 \right.
\end{eqnarray}
where, $E\left[\mathrm n_\mathrm r^{2\mathrm s+1}\mathrm l_\mathrm
j\right]$ represents the eigen-values of the radial equations
(\ref{radl.eqn}) with various quantum numbers $\mathrm n_\mathrm r$,
j, l, and s. Because the parameters $a$, $b$, $c$ and $d$ are
involved in the potential (\ref{v-s}), $E\left[\mathrm n_\mathrm
r^{2\mathrm s+1}\mathrm l_\mathrm j\right]$ must be functions of
these parameters. $m[\mathrm{meson}]$ are the masses of the
individual states which are shown in the following table
\cite{Amsler:2008zzb}:

\begin{center}
\begin{table}[!h]
\caption{The experimental central values of the spectra of the
$c\bar{c}$ charmonia states} \label{tab1}
\begin{tabular}{c|c|c|c|c|c}
 \toprule[1pt]
  state  & m(GeV)  & state &
  m(GeV)& state & m(GeV)
  \\ \midrule[0.5pt]
  $J/\psi(1^3\mathrm S_1)$ & 3.0969 &
    $\chi_{c1}(1^3\mathrm P_1)$ & 3.5107 &
    $\psi(2^3\mathrm S_1)$ & 3.6861  \\
     $\chi_{c0}(1^3\mathrm P_0)$ & 3.4148 &
    $\eta_c(2^1\mathrm S_0)$ & 3.6370 & &\\ \bottomrule[1pt]
\end{tabular}
\end{table}
\end{center}

Sequentially, the parameters $a$, $b$, $c$ and $d$ are obtained by
solving Eqs.(\ref{e-g}). By means of the Newton's iterative method,
we have achieved as (The details about the numerical method can be
found in Ref. \cite{Press:2007}.)
\begin{eqnarray}\label{abcd}
 a=1.1715,~b=1.2250,~c=0.8087,~d=0.5291
\end{eqnarray}

Here we set $\alpha_s=0.36$ and $\kappa=0.42$ GeV which seem somehow
different from the values given in
literature\cite{Ke:2009sk,Ding:1988,Bali:1992}. But as noticed, the
deviation may be included in the phenomenological parameters $a$,
$b$, $c$ and $d$.

A few words are about our choice of the input. In principle, any
five well measured states of the charmonia can be used as the input.
However, unfortunately, the relationship between the $E\left[\mathrm
n_\mathrm r^{2\mathrm s+1}\mathrm l_\mathrm j\right]$ and the
parameters in (\ref{e-g}) is complicated,  taking the central values
of the masses of $J/\psi$, $\chi_{c0}(1\mathrm P)$,
$\chi_{c1}(1\mathrm P)$, $\eta_c(2\mathrm S)$ and $\psi(2\mathrm S)$
as the inputs one can obtain reasonable solutions, otherwise, the
equations (\ref{e-g}) do render solutions for $a$, $b$, $c$ and $d$.
The reason is due to the experimental errors.

Given  $a$, $b$, $c$ and $d$ in (\ref{abcd}), the masses of the
charmonia states can be written as:

\begin{eqnarray}\label{f-r}
 M_1(\mathrm n_\mathrm r^{2\mathrm s+1}\mathrm l_\mathrm j)=E\left[
  \mathrm n_\mathrm r^{2\mathrm s+1}\mathrm l_\mathrm j\right]+E_0
\end{eqnarray}
where, $E_0$ is the zero-point energy:
\begin{eqnarray}\label{zpe}
 E_0=m[J/\psi]-E[1^3\mathrm S_1]
\end{eqnarray}
and the final result is shown in Table-\ref{tab2} below:

\begin{center}
\begin{table}[!h]
\caption{The mass spectra for the charmonia states
 (in GeV), with $m_c=1.84\mathrm {GeV}$.
 The mass of the EXP is the value given in PDG\cite{Amsler:2008zzb}.} \label{tab2}
\begin{tabular}{lcc|lcc}
 \toprule[1pt]
  meson &
   EXP &
    Prediction &
     meson & EXP & Prediction
    \\ \midrule[0.5pt]
    $\eta_c(1^1\mathrm{S}_0)$ & 2.9803 & 3.0189 &
     $\chi_{c2}(1^3\mathrm{P}_2)$ & 3.5562 & 3.5564
    \\
    $J/\psi(1^3\mathrm{S}_1)^{\mathrm{fit}}$ & 3.0969 & 3.0969 &
     $\eta_c(2^1\mathrm{S}_0)^{\mathrm{fit}}$ & 3.6370 & 3.6370
    \\
    $\chi_{c0}(1^3\mathrm{P}_0)^{\mathrm{fit}}$ & 3.4148 & 3.4148 &
     $\psi(2^3\mathrm{S}_1)^{\mathrm{fit}}$ & 3.6861 & 3.6861
    \\
    $\chi_{c1}(1^3\mathrm{P}_1)^{\mathrm{fit}}$ & 3.5107 & 3.5107 &
     $\psi(3^3\mathrm{S}_1)$ & / & 4.1164
    \\
    $h_c(1^1\mathrm{P}_1)$ & 3.5259 & 3.5100 & & &\\
     \bottomrule[1pt]
\end{tabular}
\end{table}
\end{center}

Explicitly, in the process, the masses of $J/\psi,
\chi_{c0},\psi(2^3\mathrm{S}_1)$ and $\chi_{c1}(1^3\mathrm{P}_1)$
are taken as inputs to obtain the parameters and then the masses of
other states in the family: $\eta_c(1\mathrm S)$, $h_c(1\mathrm P)$,
$\chi_{c2}(1\mathrm P)$
and $ \psi(3\mathrm S) $. The numbers are predicted.\\

\section{The mass spectrum as the Lamb
shift is taken into account}\label{sec-4}

As well known, the Lamb shift is due to the vacuum fluctuation and
may cause sizable effects on the meson spectra. Indeed, the QED Lamb
shift may not be very significant because of smallness of the fine
structure constant $\alpha$, but for the QCD case, the situation
will be different.

On the other hand, the Breit-Fermi Hamiltonian used in
Section-\ref{sec-3} does not include the effects of the Lamb shift
in the eigen-energy (\ref{radl.eqn}). In this section, we will take
the Lamb shift into account. However, we do not introduce the
Hamiltonian induced by the Lamb shift into the differential equation
because the corresponding pieces are very complicated and it is not
necessary to do so. Instead, we simply add the estimated values of
the effects to the binding energies of various states. Repeating the
procedure done in last sections and adding the Lamb shift effects to
the spectra, we re-fit the data to obtain $a$ $b$ $c$ and $d$ again
and predict the mass spectra of the rest resonances.

Namely, we set the mass of a bound state as:
\begin{equation}
2m_c+E+\Delta E_\mathrm{LS}=M^\mathrm{EXP},
\end{equation}
where $E$ is the solution of the eigen-equation, $\Delta
E_\mathrm{LS}$ is the energy caused by the Lamb shift. Solving the
equation, one can obtain the parameters again.

The authors of Ref.\cite{Hoang:2001rr,Titard:1993nn} gave the
theoretical expressions for the binding energies  which involve
contributions of the Lamb shift. When we only concern the Lamb
shift, we must single it out from the general formulas. It is not
difficult, as a matter of fact, because the Lamb Shift starts at
$O(\alpha_s^3)$\cite{Titard:1995}. The Lamb Shift can be written as:
\begin{subequations}\label{ls}
\begin{eqnarray}
 \Delta E[\mathrm n,\mathrm j,\mathrm l,\mathrm s]&=&m\left[\Delta E(\alpha_s^3)
 +\Delta E(\alpha_s^4)+\Delta
 E(\alpha_s^5)\right.\nonumber\\
 & &+\left.
 \Delta
 E(\alpha_s^6)+\dots\right]
\end{eqnarray}

For readers' convenience, let us directly copy Titard's formulas
\cite{Titard:1993nn} below, where we dropped the tree-level terms
and the relativistic corrections, and we have:
\begin{widetext}
\begin{eqnarray}
 \Delta E(\alpha_s^3)&=&-\alpha_s^3{C_F^2\over8\pi
 n^2}\left(2\beta_0\gamma_E+4a_1\right);\\
 \Delta
 E(\alpha_s^4)&=&-\alpha_s^4{C_F^2\over4n^2\pi^2}\left\{(a_1
 +\gamma_E{\beta_0\over2})^2
 +2\left[\gamma_E(a_1\beta_0
 +{\beta_1\over8})+({\pi^2\over12}
 +\gamma_E^2){\beta_0^2\over4}+b_1\right]\right\}.
\end{eqnarray}
\end{widetext}
Hoang et al.  estimated the contribution of higher orders
$O(\alpha_s^5)$ and $O(\alpha_s^6)$ to the binding
energies\cite{Hoang:2001rr}. Phenomenologically, these high-order
terms can be attributed to the effects of the Lamb Shift:
\begin{widetext}
 \begin{eqnarray}
  \Delta
  E(\alpha_s^5)&=&\alpha_s^5\log\alpha_s{C_F^2\over4\pi n^2}
   \left\{{C_A\over3}\left[{C_A^2\over2}
    +{4C_AC_F\over\mathrm n(2\mathrm l+1)}+{2C_F^2\over\mathrm n}\Big({8\over2\mathrm l+1}
     -{1\over\mathrm n}\Big)\right]+{3\delta_{\mathrm l0}C_F^2\over2\mathrm n}
      (C_A+2C_F)\right.\nonumber\\
  & &-\left.{7C_AC_F^2\delta_{\mathrm l0}\delta_{\mathrm s1}\over3\mathrm n}
   -{C_AC_F^2(1-\delta_{\mathrm l0}\delta_{\mathrm s1})\over4\mathrm n\mathrm l
    (\mathrm l+1)(2\mathrm l+1)}
     (4X_{\mathrm l\mathrm j\mathrm s}
      +<S_{12}>_{\mathrm l\mathrm j\mathrm s})\right\};\\
  \Delta
  E(\alpha_s^6)&=&\alpha_s^6\log^2\alpha_s{C_F^2\over4\pi^2\mathrm n^2}\left\{
   {\delta_{\mathrm l0}C_F^2\over6\mathrm n}
    \left[\beta_0({13C_A\over2}-C_F)+{C_A\over3}(25C_A+22C_F)\right]
     -{C_AC_F^2\delta_{\mathrm l0}\delta_{\mathrm s1}\over6n}
      \left[5\beta_0+7C_A\right]\right.\nonumber\\
  & &-\left.{C_AC_F^2(1-\delta_{\mathrm l0})\delta_{s1}
   \over8\mathrm n\mathrm l(\mathrm l+1)(2\mathrm l+1)}
    \left[\beta_0\right(2X_{\mathrm l\mathrm j\mathrm s}
     +{1\over2}<S_{12}>_{\mathrm l\mathrm j\mathrm s}\left)
      +C_A(2X_{\mathrm l\mathrm j\mathrm s}
       +<S_{12}>_{\mathrm l\mathrm j\mathrm s})\right]\right\}.
 \end{eqnarray}
\end{widetext}
\end{subequations}
$\mathrm n$ in Ref.(\ref{ls}) stands for the principal quantum
number as $\mathrm n=\mathrm n_\mathrm r+\mathrm l$, where, $\mathrm
n_\mathrm r$ and $\mathrm l$ are defined in Section-\ref{sec-3}. All
the constants as $a_1$, $a_2$, $b_1$, $\beta_{i}\; (i=1,2,3)$ are
given in Ref.\cite{Pineda:1998} (also see
\cite{Billoire:1979ih,Fischler:1977yf,Titard:1993nn,Peter:1996ig,Schroder:1998vy}).

The Lamb Shift $\Delta E[\mathrm n,\mathrm j,\mathrm l,\mathrm s]$
depends on the coupling constant $\alpha_s$ (see Eq.(\ref{ls}))
\cite{Titard:1993nn} as:
\begin{widetext}
\begin{eqnarray}
 \alpha_s(\mu^2)={2\pi\over\beta_0\ln \mu/\Lambda}\left\{
  1-{\beta_1\over\beta_0^2}{\ln(\ln\mu^2/\Lambda^2)\over\ln\mu^2/\Lambda^2}
  +{\beta_1^2\ln^2(\ln\mu^2/\Lambda^2)-\beta_1^2\ln(\ln\mu^2/\Lambda^2)
   -\beta_1^2+\beta_2\beta_0\over\beta_0^4\ln^2\mu^2/\Lambda^2}\right\}.
\end{eqnarray}
\end{widetext}
It is noted that  unlike the others in the full Hamiltonian which
can be written in the pure operator form, the contributions of the
Lamb shift to the spectrum energies are always associated with the
concrete states.

Using the formulas given above, one can evaluate the Lamb Shift of
the charmonia states. The scheme of renormalization is suggested by
Pineda et al.\cite{Titard:1993nn,Pineda:1998}. Actually, there is a
term $\ln\left[\mathrm na(\mu^2)\mu\over2\right]$ in the theoretical
expression of the energy (see \cite{Titard:1993nn,Hoang:2001rr}),
where, $a(\mu^2)$ stands for the Bohr radius and $\mu$ is the
renormalization scale:
\begin{subequations}
\begin{eqnarray}
 a(\mu^2)={2\over mC_F\tilde{\alpha}_s(\mu^2)}
\end{eqnarray}
where,
\begin{eqnarray}
 \tilde{\alpha}_s(\mu^2)&=&\alpha_s\left\{1
  +\left(a_1+{\gamma_E\beta_0\over2}\right){\alpha_s\over\pi}
   \left[\gamma_E\left(a_1\beta_0+{\beta_1\over8}\right)\right.\right.
  \nonumber\\
 & &+\left.\left.({\pi^2\over12}+\gamma_E^2){\beta_0^2\over4}+b_1\right]
     {\alpha_s^2\over\pi^2}\right\}.
\end{eqnarray}
\end{subequations}
If one defines  \cite{Pineda:1998}:
\begin{eqnarray}
 \mu={2\over\mathrm na}
\end{eqnarray}
this choice of $\mu$ will cancel the terms related to $\ln[{\mathrm
na\mu\over2}]$ in the spectrum energy.

The value of the parameter $\Lambda$ is near 0.30
GeV\cite{Pineda:1998}. Here, we choose it as 0.275 GeV. The reason
is that, at this point, $\alpha_s^{\mathrm n=2}=0.38$, just near to
the value of $\alpha_s$ we used in Section-\ref{sec-3}.

It is obviously different from the conventional renormalization
scheme we commonly used. A consequence is that the coupling constant
$\alpha_s$ is different for different quantum number n:
\begin{eqnarray}\label{as}
 \alpha_s^{\mathrm n=1}=0.31~,~\alpha_s^{\mathrm n=2}=0.38~
  ,~\alpha_s^{\mathrm n=3}=0.43
\end{eqnarray}

Simply adding the Lamb shift to the total binding energy is like
that we change the zero-point energy for each state. We still select
masses of $J/\psi$, $\chi_{c0}(1\mathrm P)$, $\chi_{c1}(1\mathrm
P)$, $\eta_c(2\mathrm S)$ and $\psi(2\mathrm S)$ as inputs, and
solve the equation (\ref{e-g}) again as what we did in the last
Section. But the value of $\alpha_s$ in (\ref{e-g}) is taken as that
value given in Eq.(\ref{as}) which depends on  $n$. The new
solutions of $a$, $b$, $c$, and $d$ are:
\begin{eqnarray}
 &&\left\{
  \begin{array}{l}
   a^{(1)}=1.3943,~b^{(1)}=1.4057,\\
   c^{(1)}=0.6243,~d^{(1)}=0.9910;\\
  \end{array}
 \right.\label{abcd-1}\\
 &&\left\{
  \begin{array}{l}
   a^{(2)}=1.4191,~b^{(2)}=1.3292,\\
   c^{(2)}=0.6459,~d^{(2)}=0.9438.\\
  \end{array}
 \right.\label{abcd-2}
\end{eqnarray}
where, expression (\ref{abcd-1}) is the solution when the Lamb Shift
is taken up to order $O(\alpha_s^3)$:
\begin{eqnarray*}
\Delta E[\mathrm n,\mathrm j,\mathrm l,\mathrm s]=m\left[\Delta
E(\alpha_s^3)\right]
\end{eqnarray*}
and (\ref{abcd-2}) is for the Lamb Shift:
\begin{eqnarray*}
 \Delta E[\mathrm n,\mathrm j,\mathrm l,\mathrm s]=m\left[\Delta E(\alpha_s^3)+\Delta
 E(\alpha_s^4) +\Delta E(\alpha_s^5)+\Delta E(\alpha_s^6)\right]
\end{eqnarray*}
up to the order $O(\alpha_s^6)$. With these two solutions, our
predictions are given in  Table-(\ref{tab3})).

\begin{center}
\begin{table}[!h]
\caption{The mass spectrum with the Lamb Shift (in GeV), where, the
LS stands for the contribution of the Lamb Shift, $M_{1,2}$ is the
predicted mass when the parameter is set as in Eq.(\ref{abcd-1}) or
Eq.(\ref{abcd-2}) and $M'_{1,2}$ stands for $M'_{1,2}=M_{1,2}+\Delta
E^{(1,2)}$.} \label{tab3}
\begin{tabular}{lcccccc}
     \toprule[1pt]
  meson & $\Delta E^{(1)}$  &
   $M_2$  &
     $M'_2$  &
      $\Delta E^{(2)}$ & $M_3$ &
       $M'_3$
    \\ \midrule[0.5pt]
    $\eta_c(1^1\mbox{S}_0)$ &
     -0.0674 & 3.0820 & 3.0146 &
      -0.1196 & 3.1612 & 3.0416
    \\
    $J/\psi(1^3\mbox{S}_1)$ &
     -0.0674 & 3.1643 & 3.0969 &
      -0.1245 & 3.2215 & 3.0969
    \\
    $\chi_{c0}(1^3\mbox{P}_0)$ &
     -0.0310 & 3.4458 & 3.4148 &
      -0.0799 & 3.4946 & 3.4148
    \\
    $\chi_{c1}(1^3\mbox{P}_1)$ &
     -0.0310 & 3.5417 & 3.5107 &
      -0.0802 & 3.5909 & 3.5107
    \\
    $h_c(1^1\mbox{P}_1)$ &
     -0.0310 & 3.5593 & 3.5283 &
      -0.0803 & 3.6063 & 3.5260
    \\
    $\chi_{c2}(1^3\mbox{P}_2)$ &
     -0.0310 & 3.6079 & 3.5769 &
      -0.0804 & 3.6552 & 3.5748
    \\
    $\eta_c(2^1\mbox{S}_0)$ &
     -0.0310 & 3.6680 & 3.6370 &
      -0.0714 & 3.7084 & 3.637
    \\
    $\psi(2^3\mbox{S}_1)$ &
     -0.0310 & 3.7171 & 3.6861 &
      -0.0728 & 3.7589 & 3.6861
    \\
    $\psi(3^3\mbox{S}_1)$ &
     -0.020 & 4.1460 & 4.126 &
      -0.0531 & 4.1746 & 4.1215
    \\
     \bottomrule[1pt]
\end{tabular}
\end{table}
\end{center}

\section{Conclusion and discussion}

In this work, we study the role of scalar potential to the spectra
of charmonia. Our strategy is that the scalar and vector potentials
have different fractions which manifest in their coefficients (in
the text, they are $a$, $b$, $c$ and $d$ for the Coulomb and
confinement pieces respectively). By fitting some members of the
charmonia family, we can fit them. Then with the obtained
parameters, we further predict the mass spectra of the rest
resonances of charmonia. It is shown that unlike the QED case where
the fraction of scalar potential is very small and negligible, the
fraction of scalar potential is of the same order of magnitude as
the vector potential. This is consistent with the conclusion of
Ref.\cite{Franklin:1998kd} and this is not surprising. As we
indicated that for the vector-like coupling theories QED and QCD,
the scalar potential can only appear at loop level or is induced by
non-perturbative effect (QCD only). Thus it is loop-suppressed.
However, for QCD, the coupling is sizable and the non-perturbative
effects somehow are significant, so one can expect the fraction of
scalar potential is large.

Moreover, the Lamb shift is induced by the vacuum fluctuation and
only appear at loop level, indeed the leading contribution is at
$O(\alpha_s^3)$. Therefore for the QED case, it is hard to observe
the Lamb shift (observation of the Lamb shift is a great success for
theory and experiment indeed), however, for QCD the effects are not
ignorable. By taking into account the Lamb shift, we re-fit the
model parameters and find they are obviously distinct from that
without considering the Lamb shift.

In the work, by studying the charmonia spectra we investigate the
contribution of higher orders of $\alpha_s$ and non-perturbative QCD
effects. However to distinguish between them, one needs to do more
theoretical researches. This result helps us to get a better
understanding of QCD, especially the non-perturbative effects. Even
though it is only half-quantitative, it is an insight to the whole
picture.

When we take into account the contribution of the Lamb shift to the
mass spectra, it is more obvious that higher order effects are
important in QCD. Because the Lamb shift only appears at order
$O(\alpha_s^3)$, its existence manifests higher order effects. Our
calculations show that while higher orders up to $O(\alpha_s^6)$ are
involved, the fitted values of a,b,c,d are different from those when
only $O(\alpha_s^4)$ is considered.

The same strategy can be applied to the bottomonia family and even
the $B_c$ resonances where one can further test the theoretical
framework and investigate the higher order QCD behaviors. That would
be the contents of our next work.

\section*{Acknowledgments}
This project is supported by the National Natural Science Foundation
of China (NSFC) under Contracts No. 10775073;  the Special Grant for
the Ph.D. program of Ministry of Eduction of P.R. China No.
20070055037; the Special Grant for New Faculty from Tianjin
University.

\end{document}